# OF THE BLACK HOLE THERMODYNAMICS

## The Concepts of "Area Entropy", "Temperature" and "Radiation"

Xinyong Fu, Shanghai Jiao Tong University, xyfu@sjtu.edu.cn

## Introduction

In early seventies of last century, Bekenstein and Hawking put forward the concepts of the "area entropy", "gravitational temperature" and "thermal radiation" of a black hole, and thus established a new branch of astrophysics, namely, the black hole thermodynamics [1] [2].

In his arxiv paper *THE ORIGIN OF ENERGY FOR THE BIG BANG* [3], the author of this paper and his cooperator Zitao Fu put forward a new model of the universe: The universe is gravitationally closed, filled up with an equilibrium radiation at a temperature of about 3 K; its main part is just such a vast heat ocean, and all the real matter is only a small fraction, which exploding and concentrating within the heat ocean repeatedly. They analyzed the process in which a black hole attracts thermal radiation from the cosmic space, asserting that, in physics, such an energy-concentrating process is very special, which reduces entropy, violating the second law of thermodynamics. They emphasized that black holes play a very special and crucial role in the great circulation of matter and energy in the universe.

The author's ideas are completely different from those of Bekenstein and Hawking. These differences are discussed in detail in this paper.

## 1. Is the area of event horizon of a black hole also entropy?

Bekenstein and Hawking hold: the second law of thermodynamics is valid everywhere; though the absorption of heat radiation from the cosmic space by a black hole reduces the entropy of the space, in the meantime, the area of the event horizon of the black hole increases; they argue that the area of the event horizon is also entropy, and the overall change in entropy in the whole absorption process is still an increase.

Contrary to their opinion, the author of this paper hold: the reasons with which Bekenstein and Hawking regard a black hole's area as entropy are inadequate; a black hole's absorbing heat from cosmic space results in a decrease of entropy, in contradiction to the second law of thermodynamics.

Let us compare these two opposite opinions and see which one is correct.

Hawking verified that the area of event horizon of a black hole never decreases. That is his famous theorem of non-decreasing of a black hole's area. Furthermore, he and Bekenstein thought that the non-decreasing behavior of a black hole's area is very reminiscent of the behavior of the entropy of a solitary system, which never decreases, either. This reminiscence led them to think that a black hole's area is a measure of its entropy.

The author holds that the "reminiscence" or "similarity" Bekenstein and Hawking speak of here is actually a false one.

Why?

As is well known, the increase of entropy of a solitary system is due to some irreversible processes happened within the system, and it has nothing to do with the exterior world, the environment.

Contrarily, the increase of a black hole's area is due to the import of matter or energy from the environment. If there is no import of matter or energy from the exterior, no matter what processes happen within the black hole, it is impossible for its area of event horizon to increase.

So, although both of the two processes are "increase of something", one is totally caused by some interior events, independent from the exterior world, and the other happens at the cost of the loss of matter and energy of the exterior world, relying nothing on any interior events. Obviously, the patterns of the two events are absolutely different, not similar at all.

Furthermore, let's make a compare between the physical mechanisms of the two processes of "increase of something".

The physical mechanism of the increase of entropy of a solitary system is some irreversible processes within the system. There are a great variety of such irreversible processes, but all of them belong to the category of thermodynamics, showing strong thermodynamic color or style, such as work converts into heat through friction or collision, heat transmits from higher temperature to lower temperature, non-uniform in pressure or density changes automatically toward uniform, etc.

On the other hand, the increase of the area of an event horizon of a black hole is simply determined by the import of matter and energy from the exterior. The relation between the increase of the area of the event horizon and the import amount of matter and energy can be determined directly and easily by the formula of the area of the event horizon, which, for a Schwarzschild black hole, is

$$A = 4\pi R^2 = \frac{16\pi^2 G^2 M^2}{c^4}.$$

Such a formula belongs purely to the category of static mechanics of relativity, and it has nothing to do with thermodynamics.

So, the physical mechanisms of the two kinds of "increase of something" are also not similar to each other at all.

Hence, there is no any real similarity between the increase of a black hole's area and the increase of the entropy of a solitary system. The comparison between these two processes made by Bekenstein and Hawking is a careless done job, and we cannot accept their hasty and rough conclusion that a black hole's area is also entropy.

Further more, in thermodynamics, there are a variety of forms of entropies, and entropies in different forms equivalent to each other quantitatively. In other words, entropy is actually an equivalent quantity. The equivalent relations between different forms of entropies are based on the reversible processes that link up the different forms of entropies. Entropy is actually conserved in any reversible process. A certain quantity of entropy in one form can convert totally through some reversible process into a certain quantity of entropy in another form, without producing any other effects. The meaning of "reversible process" here is well known: the finally obtained entropy can convert back totally through a process carried out in the opposite direction step by step into the original entropy, without producing any other effect. By the way, a reversible process can also be expressed in a broader way as follows: after it has occurred, both the system and the environment can restore to their original states entirely through whatever a process. Hence, if there is someone who intends to introduce a new entropy into physics, he or she should find out such and such a reversible process that can link up the new entropy with any of the traditional entropies, for example, the heat-temperature-quotient entropy or volume entropy, and determine the special quantitative equivalent relation between the new one and the traditional one.

Can we find any reversible process that can link up a black hole's area with some traditional entropy? The process of thermal radiation's falling into a black hole is an extremely violent spontaneous process, and we cannot control it in any way. Such a process itself is certainly an irreversible process. Neither can we find any other reversible process that can link up the above mentioned two quantities and determine their quantitative relation. Hence, there is no reason to regard a black hole's area as entropy.

Bekenstein, Hawking and others tried again to use Boltzmann's relation $S = k \ln \Omega$ to show that a black hole's area is entropy [4] [5]. Nevertheless, everyone knows that $\Omega$ is the number of the microscopic states corresponding to a statistic distribution. The physical state of a black hole, according to Bekenstein, Hawking and some others, can be completely determined by three quantities, namely, its mass $M$, angular speed $\omega$ and electric charge $Q$. Therefore, the general number of the microscopic states of a black hole is only one. According to Boltzmann's relation, its entropy is zero. Still they won't give up, they invented here a fantastic explanation that $\Omega$ is " the total number of the microscopic ways of the formations of the black hole ".

Entropy is a state quantity of a system. The entropy of a system at a certain instant is determined only by the state at the instant, and it has nothing to do with its past history. To speak in more conventional words in thermodynamics, the entropy of a system at a certain state has nothing to do with the special process or path through which the system reached the state. And of course, this state quantity is even more impossible to be related to the number of the possible microscopic ways through which the system can reach its present state!

Thermodynamics should not be "developed" in such a distorted way.

Let's proceed further.

It is rather comical that Hawking verified later that a black hole has radiation. It radiates thermal radiation as well as matter particles. These radiations, just as he said, result in the reduction of the mass of the black hole, and hence the reduction of the area of the black hole. So, a black hole's area can also decrease. And then, is the theorem of "no-decreasing of a black hole's area" still valid? As the theorem is no longer valid, how can they still speak of the "similarity" between a black hole's area and the entropy of a solitary system? Everyone knows that the entropy of a solitary system never decrease, which is valid in all the cases.

The author's opinion is as follows. The event horizon of a black hole permits thermal radiation passing through it from the exterior into the black hole, never from the interior to the exterior space, so it is a "one-way membrane" for thermal radiation. In other words, it is a Maxwell's demon in nature. Stars give off light and heat, that is energy going from a concentric state to a scattered state, from a high temperature to a low temperature, resulting in the increase of entropy; black holes, quite contrarily, collect thermal radiation from the space, that is energy going from a scattered state to a concentric state, from a low temperature to a high temperature, resulting in the decrease of entropy. It is very apparent that the two kinds of processes are opposite to each other, and from the viewpoint of energy circulation, they are actually complementary to each other. The author wonders that why these people cannot catch sight of so simple and so explicit a fact.

Bekenstein and Hawking may argue that a black hole's "radiation" is always connected to a "temperature" (the gravitational acceleration at the event horizon), and the quotient of the radiated heat from a black hole to its "temperature" is also entropy. So, in a Hawking radiation, the area entropy is actually transformed into heat-temperature entropy and flows outward into the space, and the overall change in entropy for the process is still not a decreasing one, in accord with the demands of the second law.

Such an argument is also incorrect. In the third part of this paper, the author will show in details that Hawking radiation is unidirectional, different from the isotropic radiation of a black body at certain temperature. Like a laser beam, due to its uni-direction, Hawking radiation can easily be concentrated and result in very high temperature. Hence the temperature of a Hawking radiation is not $T = \kappa/2\pi k$, but extremely high. Hence, the outward flow of heat-temperature entropy in a Hawking radiation is about zero. In short, there is no area entropy that changes into heat-temperature entropy and flows outward from the black hole to the space.

Taking Hawking radiation into account, the theorem of non-decreasing of a black hole's area is no longer valid, and Bekenstein and Hawking's idea that a black hole's area means entropy is no longer valid, either.

## 2. Geroch's "gravitational engine" and a black hole's "temperature"

Via a detailed discussion on the cycle of an ideal "gravitational engine", Geroch and Bekenstein proved that the gravitational acceleration near the event horizon of a black hole corresponds to a certain "temperature", the relation between the two is $T = \kappa/2\pi k$.[6]

The author holds that their discussion is wrong.

The "gravitational engine" they described is a cycling device. Like all the previous heat engines, the device works between a high temperature heat reservoir and a low temperature heat reservoir. Firstly, how heat is spent and how work is done in a cycle is described. In their succeeding discussion, a quantum mechanical restriction to the size of a box, which is used to carry thermal radiation, is cited. Wien's displacement law of black body radiation is cited. The famous Carnot's principle is also cited, giving an upper limit to the efficiency of heat-work conversion. And so on. Finally, through a comparison between two inequalities, the relation between the gravitational acceleration at the event horizon and the black hole's "temperature", $T = \kappa/2\pi k$, is obtained. The whole verification involves so many important concepts of classical and modern physics, and sets forth a series of smart mathematical calculates, hence it looks very profound and integrated.

Nevertheless, there is one thing astonishing in their discussion: In each cycle, the heat extracted from the high temperature is equal to the heat released to the low temperature! That is to say, no any net heat changes into work. Then, where does the work performed come from? According to them, all the work done by the "engine" comes from the reduction of the gravitational potential energy of the heat radiation, which, according to relativity, has mass, too. How much this potential energy is decreased, how much work is done. Nevertheless, everyone knows that, in physics, gravitational potential energy can be used directly to do work, and this has nothing to do with heat-work conversion (i.e., heat engine). So, it is clear that Geroch's device is totally not a heat engine, but a simple mechanical apparatus. How ridiculous it is to apply Carnot's principle to an apparatus that is totally not a heat engine! The whole proof is just knocking things together to get the pre-selected conclusion, $T = \kappa/2\pi k$.

It seems that Geroch and Bekenstein, in their discussion, went backwards to James Watt and Sadi Carnot's time. Most people of the time believed in caloric theory (Carnot believed in caloric theory when he was young). According to that theory, a steam engine extracts a certain amount of heat from a high temperature in each cycle, after doing work, all the heat is released to the low temperature. There is no net decrease of the heat. It was impossible for people of that time to think that the total amount of the heat would decrease, as they believed that "the caloric itself is conserved." They just thought that in a steam engine work is performed by the driving force of the flow of the caloric from the high temperature to the low temperature. Such a "mechanism" is very similar to the one of a water turbine in a hydraulic power plant. As is well known, a water turbine relies only on the driving force of the flowing water from a higher altitude to a lower one. The mass of water here is conserved. The work obtained is not at the cost of any loss of the mass of the water.

The establishment of the first law and the second law of thermodynamics refuted down the caloric theory to the last point. The mechanism of a heat engine is totally different from the one of a water turbine! A steam engine extracts certain amount of heat from the higher temperature in a cycle, a fraction of the heat is released to the low temperature, the left is converted into work, and the total amount of the heat is decreased. Heat engine is just such an apparatus that converts part of the heat into work.

In Geroch and Bekenstein's "gravitational engine", the heat extracted from the high temperature equals the heat released to the low temperature, so its mechanism is completely identical to a water turbine. Such an apparatus is certainly not a heat engine. Actually, they missed here the definition of a heat engine.

There is nothing profound or abstruse here, and it is certainly impossible that Geroch and Bekenstein really did not know these ABC of thermodynamics. Their mistake was only a careless slip. Nevertheless, there is a real cause of the careless slip, which is their endeavor to establish an intrinsic link between the gravitational acceleration and a black hole's "temperature", and such a link actually does not exist. Marching in a totally wrong direction tends to lead people falling down.

### 3. Hawking radiation, its "spectrum" and "temperature"

Hawking radiation is due to the quantum fluctuation of vacuum and the gravitational acceleration near the event horizon of a black hole. Hawking found that the mathematical expression of the spectrum of the radiation, which he derived, is pretty similar to the one of a black body radiation at certain temperature, differing only a constant factor between the gravitational acceleration of the former and the temperature of the latter, $\kappa/2\pi k = T$. Hence,

Hawking and others asserted that the gravitational acceleration here is actually the black hole's "temperature".

The author holds that, the "similarity" between the two expressions Hawking found is only some mathematical coincidence. As viewed from thermodynamics, it is incorrect to hold that the gravitational acceleration near the event horizon represents the black hole's "temperature". The reasons are as follows.

The physical mechanism of black body radiation, according to the original ideas of Rayleigh-Jeans and Planck, is the random thermal motion of numerous electro-magnetic harmonic oscillators of various wavelengths in a cavity at a thermal equilibrium state. The oscillators obey Boltzmann's distribution, and the spectrum of the radiation depends on the temperature that is determined by the intensity of the random motion of all the oscillators. If we look at it in another way, i.e., from the viewpoint of photon gas, the physical mechanism of black body radiation is the random thermal motion of numerous photons (bosons, obeying Bose-Einstein distribution), and the spectrum of the radiation depends on the temperature that is determined by the intensity of the random motion of all the photons. In a word, the physical mechanism of a black body radiation is the random thermal motion of a great number of quasi-independent subsystems in an equilibrium thermodynamic system, and the temperature is a measurement of the intensity of the random thermal motion of all the quasi-independent subsystems.

The physical mechanism of Hawking radiation is the quantum fluctuation of vacuum near the event horizon, which is related closely to the gravitational acceleration at the site. The author's opinion is: gravitational acceleration is just the intensity of the gravitational field, and that is all, nothing more. Gravitational acceleration is totally different from temperature, i.e., totally different from the intensity of the random thermal motion of numerous microscopic particles.

To understand better the difference between the gravitational acceleration and the temperature here, let us make some further comparison between the two radiations.

First, black body radiation is isotropic, while Hawking radiation is fairly unidirectional.

Black body radiation is equilibrium thermal radiation. Due to its origin of random thermal motion, isotropic is an absolutely necessary and essential characteristic of the radiation.

Hawking radiation is due to the quantum fluctuation of vacuum and the gravitational acceleration, and it is fairly unidirectional. Take the radiation in Kerr-Newman-de Sitter time-space as an example, the outward wave function of the scalar particles is

$$\Phi_{\omega}^{out} = N_{\omega} \left\{ y(r - r_{+}) \Psi_{\omega}^{out}(r - r_{+}) + y(r_{+} - r) \Psi_{\omega}^{out}(r_{+} - r) \exp\left[\frac{\pi}{k_{h}}(\omega - \omega_{0})\right] \right\},$$

and the expression of $N_{\omega}$ here is

$$N_{\omega}^{2} = \frac{1}{e^{2\pi(\omega - \omega_{0})/\kappa} - 1}.$$

Out the event horizon, the above wave function represents the photon current that is produced at the event horizon and going outward in the direction of the radius of the black hole [7]. Obviously, this is a unidirectional radiation.

What is the difference between the properties or behaviors of isotropic radiation and unidirectional radiation?

The black body radiation at certain temperature is isotropic, that means, its radiations in all the directions in the 4π solid angle are identical in spectrum as well as in intensity. No matter how we tried to concentrate such a radiation, it is impossible to form a temperature higher than the original temperature of the radiation. For example, the temperature at the surface of the sun is about 6000°C, hence the light and heat ejected by it is a black body radiation of 6000°C. With a lens or lots of mirrors, we can concentrate the sun's radiation to obtain a high temperature. Nevertheless, the highest temperature we can obtain this way is just 6000°C.

Hawking radiation is quite different. Due to its fine uni-directionality, it should be very easy to concentrate it to form an extremely high temperature. This is similar to a laser beam, which, due to its uni-directionality, is very easy to be concentrated to form a very high temperature. Therefore, as viewed from thermodynamics, the temperature of Hawking radiation is not $T = \kappa / 2\pi k$, but very high, approximately infinite.

Some people may argue that, the general wave function of Hawking radiation is

$$\Phi(t, r, \vartheta, \phi) = e^{-i\omega t} e^{im\phi} \Phi(r, \theta) = e^{-i\omega t} e^{im\phi} \Theta(\theta) R(r).$$

It seems to be also related to the angular coordinates $\theta, \phi$, and hence, it is not unidirectional, but isotropic.

This argument is not correct. The vertexes of the angles $\theta, \phi$, i.e., the origin of all the coordinates related here ($\theta, \phi$ and $r$), is the center of the black hole, not the site on the event horizon where the photons are produced. After their birth, the photons do not fly away from the birth site in all the directions. They only fly away along the radius of the black hole. So, it is a unidirectional radiation. Actually, $\theta, \phi$ just represent the coordinates of the position of the birth site of the photons on the event horizon.

Second, the energy volume density of a black body radiation in an equilibrium cavity in the infinitesimal frequency range between $\omega$ and $\omega + d\omega$ is

$$u(\omega, T)d\omega = \frac{1}{\pi^2 c^3} \frac{\hbar \omega^3}{(e^{\hbar\omega/kT} - 1)} d\omega .$$

This is Planck's formula for black body radiation. The formula shows that the spectrum of a black body radiation is determined mainly by two factors, one is the Bose-Einstein factor $(e^{\hbar\omega/kT} - 1)^{-1}$, and the other is $\hbar \omega^3$.

Let us have a look at the origin of the two factors.

(1) Bose-Einstein factor $f_s = (e^{\hbar\omega/kT} - 1)^{-1}$. This factor comes from Bose-Einstein statistical theory, and it represents the number of photons at each quantum state of the frequency $\omega$ in a photon gas at an equilibrium state.

(2) Factor $\hbar \omega^3$. $\hbar \omega^3$ consists two sub-factors, $\omega^2$ and $\hbar \omega$. (a) The origin of $\omega^2$: in a cavity of volume V, the phase-space volume of photons between $p$ and $p + dp$ is $8\pi V p^2 dp$. As $p = \varepsilon/c = \hbar\omega/c$, the number of quantum states between $\omega$ and $\omega + d\omega$ is

$$\frac{8\pi V p^2 dp}{h^3} = \frac{V}{\pi^2 c^3} \omega^2 d\omega .$$

In the above expression, the different momentums in all the directions in the $4\pi$ solid angle have been taken into account, that is, the isotropy of the black body radiation has been taken into account. (b) The factor $\hbar\omega$: as is well known, it is the energy of one photon.

Planck's formula represents the monochromatic radiation of a black body at certain temperature in the frequency range between $\omega$ and $\omega + d\omega$. Integrate it over $\omega$ from zero to infinite, we obtain the full-color radiation, which coincides with Stefan-Boltzmann's law. The result, i.e., the energy volume density of the full-color radiation is $u = \alpha T^4$, and the corresponding area radiation intensity is

$$\Gamma = \frac{1}{4} c u = \sigma T^4 .$$

The above discussion is of the spectrum and full-color radiation of black body radiation. Now let us have a look at Hawking radiation.

In a Hawking radiation, the photon current $N_\omega^2 = (e^{2\pi(\omega - \omega_0)/\kappa} - 1)^{-1}$ itself is the intensity of monochromatic radiation of the frequency $\omega$ (per unit area). It is unidirectional, that is, its photons are not radiated from their birth site in all the directions in the $4\pi$ solid angle. They are parallel beams, in the direction of the radius of the black hole. Hence, there is no longer the problem of representing the number of quantum states by the expression

$$\frac{8\pi V p^2 dp}{h^3} = \frac{V}{\pi^2 c^3} \omega^2 d\omega .$$

Hence, the factor $\omega^2$ will not appear in the spectrum expression of Hawking radiation. There are also two factors in the spectrum of Hawking radiation, the first one is $(e^{2\pi(\omega-\omega_0)/\kappa}-1)^{-1}$, similar to the Bose-Einstein factor, and the second one is $\hbar\omega$, the energy of one photon, not $\hbar\omega^3$. Hence, the spectrum of a Hawking radiation must be different from the spectrum of a black body radiation by a factor of $\omega^2$. They are different!

Consequently, due to the difference of $\omega^2$, the integration or summation of the monochromatic radiations of a Hawking radiation over $\omega$, i.e., its full-color radiation, should also be different from Stefan-Boltzmann's law. It cannot be $\Gamma = \sigma T^4$.

In short, Hawking radiation is obviously different from black body radiation.

According to Boltzmann's principle of detailed balance, a macroscopic thermal equilibrium must be the aggregate of the corresponding detailed balance. For an equilibrium thermal radiation, at any site (within the cavity) and in every two opposite directions in the $4\pi$ solid angle (the positive direction and the corresponding negative one), the radiation of every wavelength keeps balanced itself. Now that the Hawking radiation is a unidirectional one, while the black body radiation is an isotropic one, and what's the more, their spectra differ for a factor $\omega^2$, very apparently, it is impossible for the two kinds of radiations to constitute a detailed balance. Accordingly, we certainly refuse to agree that when the two radiations are at a "same temperature" ($\kappa/2\pi k = T$), they can keep equilibrium to each other. From this point of view, it is also wrong to regard Hawking radiation as a thermal radiation at certain temperature.

Let us summarize the above comparison between the two radiations. The factor $(e^{2\pi(\omega-\omega_0)/\kappa}-1)^{-1}$ in the spectrum formula of Hawking radiation is really very similar to the Bose-Einstein factor $(e^{\hbar\omega/kT}-1)^{-1}$ in the spectrum formula of a black body radiation. So it is understandable that Hawking and others imagined that the acceleration near the event horizon might actually be the black hole's "temperature". Nevertheless, black body radiation is isotropic, while Hawking radiation is unidirectional, and that is a major and crucial difference between the two radiations. Further, their spectra do not coincide to each other, differing with a factor $\omega^2$. All these differences reflect profoundly the different origin of the two radiations: the isotropic one (black body radiation) results from the random thermal motion, the unidirectional one (Hawking radiation) results from the quantum vacuum fluctuation under strong gravitational action. Hence, we conclude that we have no reason to regard the gravitational acceleration near the event horizon as the "temperature" of the black hole. The similarity between the two factors, $(e^{2\pi(\omega-\omega_0)/\kappa}-1)^{-1}$ and $(e^{\hbar\omega/kT}-1)^{-1}$, is only some mathematical coincidence.

By the way, Hawking radiation and a black hole's "temperature" sound extraordinary and attracting, nevertheless, they are actually not of great importance. According to Hawking and others' calculation, the "temperature" of a black hole is

$$T = \frac{\kappa}{2\pi k} = \frac{hc^3}{8\pi GMk} = 10^{-7} M_\Theta / M \, .$$

For black holes as large as a star, this "temperature" is lower than 0.000,001 K. The influence of the radiation corresponding to such a "temperature" on the evolution of the black hole itself is extremely small, and it can practically be neglect. Its influence on the exterior world is even more negligible. For black holes as large as a galaxy, their "temperature" and radiation are even much lower and weaker, and no one will take into account their influence to their own evolution as well as to the exterior world. As to the "primordial small black holes", whose masses are of the grade of $10^{12}$ kg, the "temperature" is high, and the radiation is powerful, even violent. Nevertheless, such "primordial small black holes" can only be formed under very particular conditions, and their high temperature and explosions are of rather short duration. Anyway, their influence on the evolution of the universe, or on the great circulation of matter and energy in the universe, is very small, not worth mentioning. Small black holes are strange and peculiar in behavior, and they are very seldom to be seen, hence, many people regard them with favor. However, in the general frame of cosmology, they are not important, they have no position.

To study and discuss the radiation and "temperature" of the black holes too much is not only useless but also detrimental. It leads people to neglect the very important role the black holes play in the evolution of the universe, namely, their function of re-concentrating matter and re-concentrating energy. Such a function is unique to black holes, and it is crucial to the great circulation of matter and energy in the universe. We should not set aside this important function of the black holes and restrain our attention to their extremely faint radiation and extremely low "temperature".

### 4. Black holes and the new thermodynamics

Beyond any doubt, thermal radiation's falling into a black hole is an energy-gathering process. Energy gathering is a completely new phenomenon in physics. It reduces entropy, violating Clausius' second law!

Stars give off light and heat into the space, scattering energy, resulting in the increase of entropy. Black holes, quite the contrary, collect the radiation from the space, gathering energy, resulting in the decrease of entropy. It is clear that the two kinds of processes are opposite each other, and they are complementary each other, too. They can be combined to form a circulation of energy in the universe.

The second law is not valid universally. It is not valid for some extraordinary processes: thermal radiation's falling into a black hole is an extraordinary process; the experiment accomplished recently by the author and his cooperator, in which thermal electrons are controlled by a magnetic demon, is also an extraordinary process [8]. The second law is valid only for ordinary processes.

Entropy increases in ordinary processes, and, it reduces in extraordinary processes. That should be an essential idea in the future thermodynamics.